\documentstyle[prb,aps,multicol]{revtex}
  \renewcommand{\narrowtext}{\begin{multicols}{2} \global\columnwidth20.5pc}
  \renewcommand{\widetext}{\end{multicols} \global\columnwidth42.5pc}

\input{psfig}

\newcommand{\f}[1]{Fig.~\ref{#1}}
\newcommand{\eq}[1]{Eq.~(\ref{#1})}
\newcommand{\eqs}[2]{Eqs.~(\ref{#1}) and~(\ref{#2})}

\def\be{\begin{equation}}
\def\ee{\end{equation}}
\def\bea{\begin{eqnarray}}
\def\eea{\end{eqnarray}}
\def\l({\left(}
\def\r){\right)}
\def\p{\protect}

\def\It{J_T}

\def\jcu{j_{cU}}
\def\jce{j_{cE}}

\begin{document}
\bibliographystyle{prsty}
\draft

\title{Comparison of flux creep and
nonlinear $E-j$ approach for analysis of
vortex motion in superconductors}

\author{D.~V. Shantsev$^{1,2}$, A. V. Bobyl$^{1,2}$,
Y.~M.~Galperin$^{1,2}$, and T.~H.~Johansen$^{1,}$\cite{0}
}
 
\address{$^1$Department of Physics, University of Oslo, P. O. Box 1048 Blindern,
0316 Oslo, Norway\\ 
$^2$A. F. Ioffe Physico-Technical Institute, Polytekhnicheskaya 26, 
St.Petersburg 194021, Russia} 
 
\maketitle
 
\begin{abstract}
Two commonly accepted approaches for simulations of thermally-activated
vortex motion in superconductors are compared. These are (i) the so-called
flux creep approach based on the expression $E=vB$ relating the electric field
$E$ to the velocity $v$ of the thermally-activated flux motion and the
local flux density $B$, and (ii) the approach employing a phenomenological
nonlinear current-voltage curve, $E(j)$. Our results show that the two
approaches give similar but also distinctly different behaviors
for the distributions of current and flux density in both a long slab and
thin strip geometry. The differences are most pronounced where the
local $B$ is small. Magneto-optical imaging of a YBa$_2$Cu$_3$O$_{7-\delta}$
thin film carrying a transport current was performed to compare the
simulations with experimental behavior. It is shown that the flux creep
approach describes the experiments far better than simulations based
on the $E(j)$ approach.
\end{abstract}

\pacs{PACS numbers:
74.25.Ha, 
74.60.Ge, 
74.76.Bz 
}

\narrowtext

\section{Introduction}

The term {\em flux creep} is used to describe
a thermally-activated motion
of flux lines in superconductors.
This motion is characterized by a velocity strongly dependent on
the local current density.
In high-temperature superconductors (HTSCs), the flux creep can be
specially pronounced because of small flux pinning energies and high
temperatures.\cite{giant,yeshurun}
An account of the flux creep is therefore
crucially important for understanding the time-dependent
magnetic behavior of HTSCs. In the literature one finds numerous
papers making use of flux creep analysis to describe the
evolution of flux and current density distributions,
current-voltage curves, magnetization and magnetic susceptibilities 
for superconductors of various
shapes.\cite{kes,burlachkov,GurBra94,Br94,Br-disk1} 
 
Interestingly, there exist today two commonly accepted approaches 
for the analysis of thermally-activated flux motion. 
The first one, the so-called \emph{flux creep} approach, assumes a
particular microscopic pinning mechanism,  
which defines the pinning energy $U$ and its
dependence on the local values of current density $j$ and flux density $B$. 
The velocity of the thermally-activated flux motion, $v$,
then determines the local electric field according to $E=vB$.
The second approach, on the other hand, employs a phenomenological 
nonlinear current-voltage relation, $E(j)$. For
brevity, we will call this the $E-j$ approach.

The present paper is devoted to a detailed comparison
of these two approaches. 
We carry out numerical simulations for the most conventional choice of 
$E(j)$ and $U(j,B)$ 
and set focus on the clear differences in the resulting behavior. 
The numerical findings are then compared to current density distributions
measured in YBaCuO films using magneto-optical imaging of flux 
density profiles. Distinct features in the observed current distributions 
allow us to conclude which approach gives the more realistic description. 
 
\section{The two approaches} 
 
To compare the two approaches we consider a one-dimensional flux creep
problem where the flux moves along the $\bf x$ direction, the magnetic
induction $\bf B$ is directed along the $\bf z$-axis, while the electric field
$\bf E$ is parallel to the $\bf y$-axis. The Maxwell equation has then
the form
\be
  \frac {\partial B}{\partial t} = - \frac {\partial E}{\partial x}\, .
\label{dBdt}
\ee
In the flux creep approach, since it represents an activation process,
the velocity of the vortex motion is given by
\be
  v = v_c {\rm e}^{-U(j,B)/kT}\,  ,
\label{E1}
\ee
where $v_c$ is the velocity when $U=0$. In the case that the pinning energy
has a logarithmic dependence on the current,
$U(j,B)=U_c \ln(\jcu/j)$, it follows that the electric field equals
\be
  E = v_c B \left( j/\jcu \right)^{U_c/kT}.
\label{E1a}
\ee

In the $E-j$ approach, the phenomenological $E(j)$
relation is usually chosen in the power law form,
\be
 E = E_c \left( j/\jce \right)^n,
\label{E2}
\ee
with $n \gg 1$, and where $\jce$ and $E_c$ are constants
with dimension of current density and electric field, respectively.

Comparing \eqs{E2}{E1a} one can see that  the
exponent $n$ in the $E-j$ approach  plays the same role as
the ratio $U_c/kT$ in the flux creep model.
However, even if $n = U_c/kT$,
there still remains an important difference.
In \eq{E1a} one has $E \propto B$,
i.e., the electric field induced by the vortex motion is
proportional to the number of moving vortices. 
In the $E-j$ approach, \eq{E2}, this proportionality is absent.
As a result, the two approaches become {\em different} if
all parameters, $\jcu$, $\jce$, and $E_c$ are {\em independent} of $B$,
which is the conventional assumption.\cite{equivalence}

In the $E-j$ approach at $n \rightarrow \infty$ the electric
field tends to zero for $j<\jce$, while it becomes infinitely large
for $j>\jce$. This situation is equivalent to the critical-state model
characterized by the critical current density $\jce$.
Similarly, for $U_c/kT \rightarrow \infty$ the flux creep approach
reproduces the critical-state model with critical current density $\jcu$.
Therefore, in the limit $U_c/kT,n \rightarrow \infty$, and
$\jcu=\jce$ both approaches become equivalent.
Accordingly, their difference is expected
to grow as $n$ and $U_c/kT$ becomes smaller.

To complete the set of equations one needs also
a relation between the flux  and current density.
Let us assume that the superconductor has infinite extension along the 
$\bf y$-axis, the direction of current, and occupies the region
$-w \le x\le w$. In the $\bf z$-direction it 
can be either infinite (a slab), or very thin (a strip) with 
thickness $d \ll w$. 
With the magnetic field $\bf{B}_a$ applied along $\bf z$,
the flux and current density can in both cases be
considered uniform in this direction. 
Making the common assumption that $B = \mu_0 H$, one has for a slab that
\be
  \mu_0 j = - \partial B / \partial x\, . 
\label{slab} 
\ee
For a thin strip the Biot-Savart law yields 
\be
  B(x) = B_a + \frac{d \mu_0}{2\pi} \int^{w}_{-w} \frac{j(u)}{u-x} \, du \, .
\label{bs} 
\ee 
It is convenient to invert the latter equation, which gives\cite{BrIn} 
\begin{eqnarray} 
  j(x) &=& \frac{2}{\pi d \mu_0} \int^{w}_{-w} \frac{B(x')-B_a}{x-x'}
  \sqrt{\frac{w^2-x'^2}{w^2-x^2}} dx'
  \nonumber \\ &&
  + \frac{I_T}{\pi d \sqrt{w^2-x^2}}\, ,
\label{strip}
\end{eqnarray}
where $I_T$ is the transport current.

In the numerical simulations we solve the set of equations
(\ref{dBdt}), (\ref{E1a}) for the flux-creep approach,
and (\ref{dBdt}), (\ref{E2}) for the $E-j$ approach, respectively. The
relation between $B$ and $j$ is taken from \eq{slab} for the case of a
slab, and from \eq{strip} in the thin strip geometry. The critical
current densities, $\jcu$ and $\jce$, are assumed to be $B$-independent.

\section{Numerical results}
\subsection{Comparison with exact solution}

As a check of the quality of our numerical simulation scheme
we compared numerical results for the $E-j$ approach
with an exact analytical solution, which can be obtained in the slab case.
Assume that at $t=0$
a finite external magnetic field is suddenly turned on. 
The flux density
distribution can then be expressed as a function of a single scaling parameter
as long as the flux fronts penetrating from opposite sides do not
overlap. With $E \propto j^n$, the scaling law has the form\cite{scal}
\be
B(x,t) = a f(\xi), \quad \quad \xi = (w-x)
a^{-\frac{n-1}{n+1}} t^{-\frac{1}{n+1}} \, .
\label{scal}
\ee
Here
\be
f(\xi) = \
\frac{ \Phi \left(\xi/\xi_0\right)}{ \Phi(0)}, \quad \quad
\Phi(z)  = \int_z^1 dx \,  (1-x^2)^{\frac{1}{n-1}},
\label{fxi}
\ee
where $\xi_0$ is given by the equation
$$
\xi_0^{n+1}\, \left[ \Phi(0) \right]^{n-1} =  2  n(n+1)/(n-1)
$$
Note that $f(\xi_0)=0$, and hence $\xi = \xi_0(n)$ describes the
advance of the flux front.
\begin{figure}
\centerline{ \psfig{figure=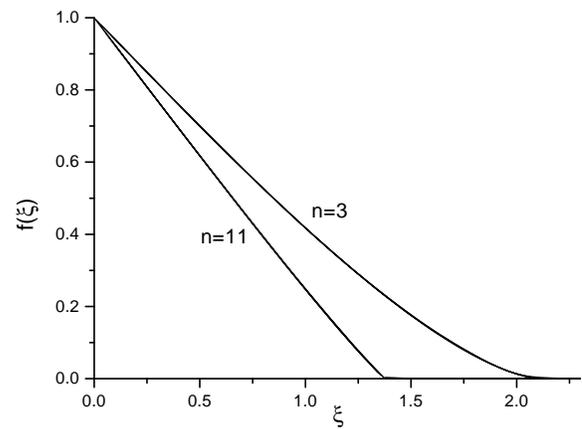,width=8cm}}
\caption{Normalized $B$-profiles obtained by numerical simulations
for a slab described by \p\eq{E2} with $n=3$ and $n=11$.
The profiles, which are plotted in the scaling variable $\xi$ 
defined in \p\eq{scal}, correspond to five different times
after a step has occured in $B_a$. The collapse among the curves 
agrees fully with our analytical solution.
\label{f_scal}}
\end{figure}

Shown in~\f{f_scal} are simulated profiles of the flux density
$B(\xi)$ in a slab using $n=3$ and $n=11$.
Both graphs contain five curves corresponding to different points
in time $t$ between 5$\tau$ and 10$^4\tau$, where $\tau=B_a^2/(\mu_0 \jce E_c)$.
Notice the Bean model like linearity in the profiles for $n=11$, and
the clear non-linearity for $n=3$.
Shown together with these curves is also the analytic solution $f(\xi)$,
given by Eq.~(\ref{fxi}). The collapse within each family of curves
demonstrates an excellent agreement, and gives confidence in
the numerical procedures.

\subsection{Slab with a transport current}

In the following we present simulation results
assuming that a transport current, linearly
increasing in time,
is passed through an initially zero-field-cooled superconductor.
The choice of parameters is $d\It/dt = 10^{-3} j_c v_c$
where $\It(t)=\int_{-w}^w j(x,t)\, dx$ is the transport current per
unit height,
$j_c \equiv j_{cU}=j_{cE}$, and $U_c/kT = n = 5$.
Moreover, we let
$E_c= 0.2 v_c \mu_0 j_c w$, which gives approximately the same average electric field
in the superconductor for both the flux creep and $E-j$ approach.

\begin{figure}
\centerline{ \psfig{figure=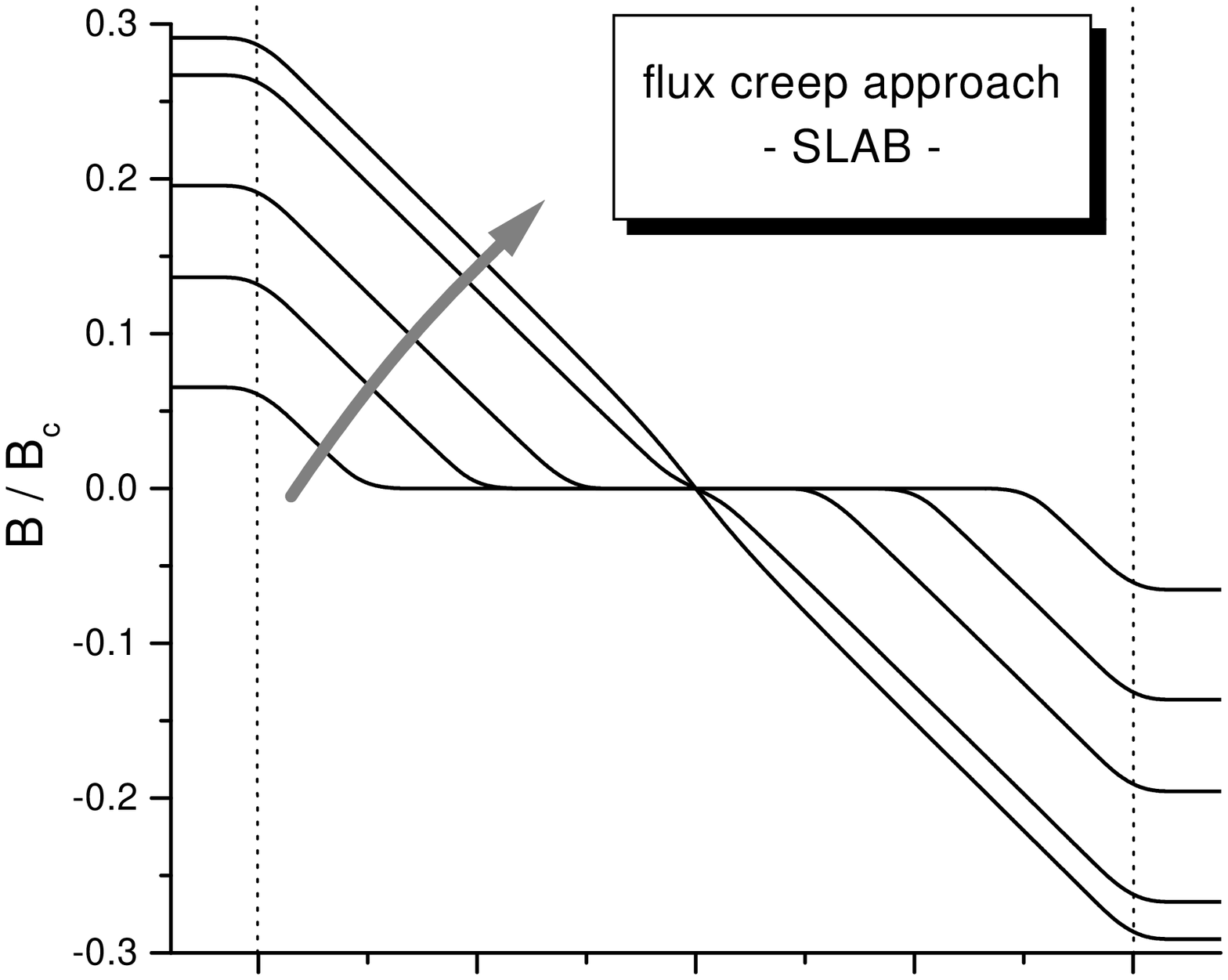,width=8cm}}
\centerline{ \psfig{figure=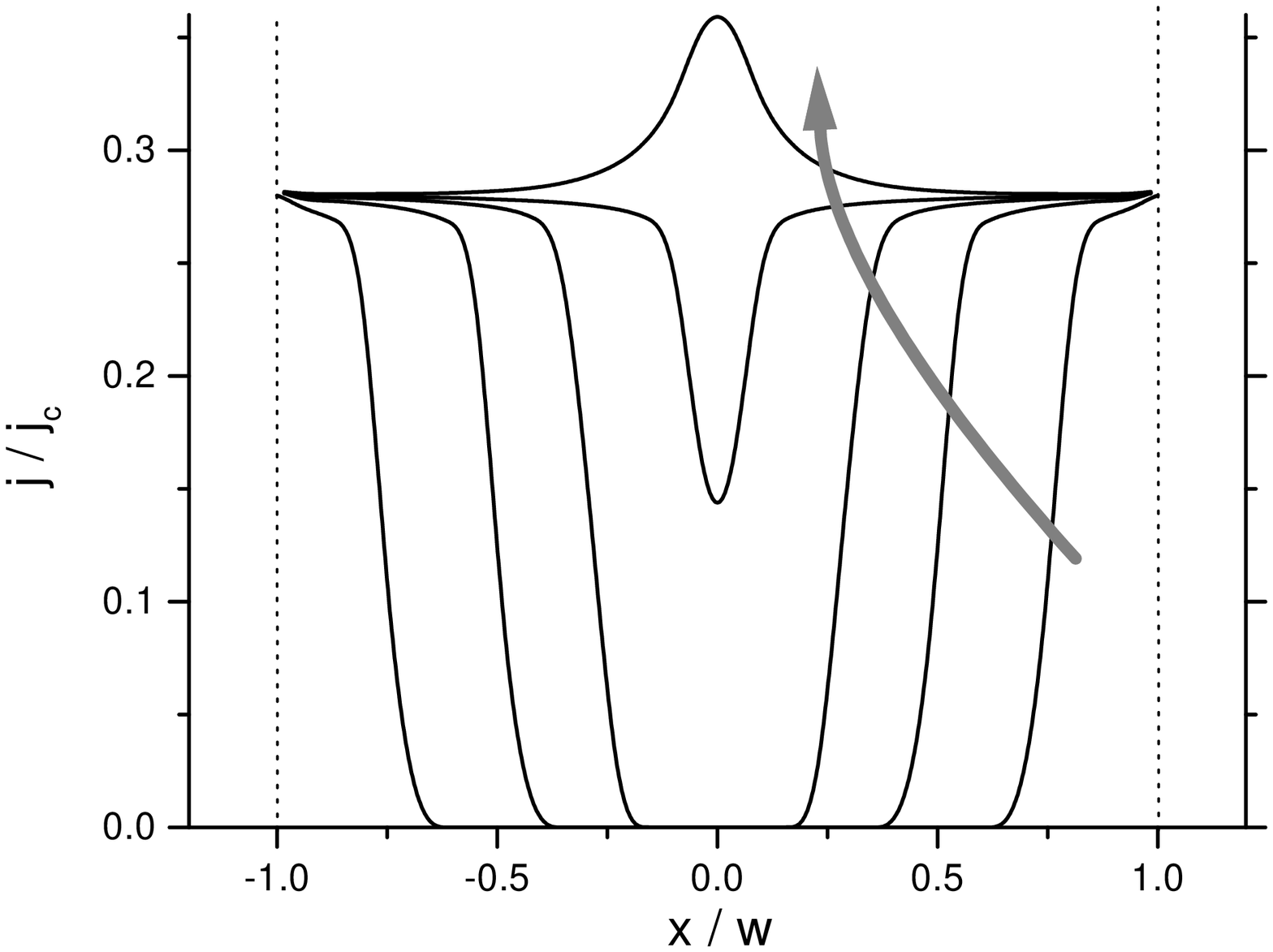,width=8cm}}
\caption{Temporal evolution of current and flux density distribution
in a slab with an increasing transport current. The graphs are obtained using the
flux creep approach. The profiles correspond to currents $\It/J_c=$ 0.07, 0.14,
0.20, 0.27, 0.29, where $J_c=2w j_c$.
The dotted line marks the edge of the superconductor.
Arrows indicate the direction of time.
\label{f_slb}}
\end{figure}

\begin{figure}
\centerline{ \psfig{figure=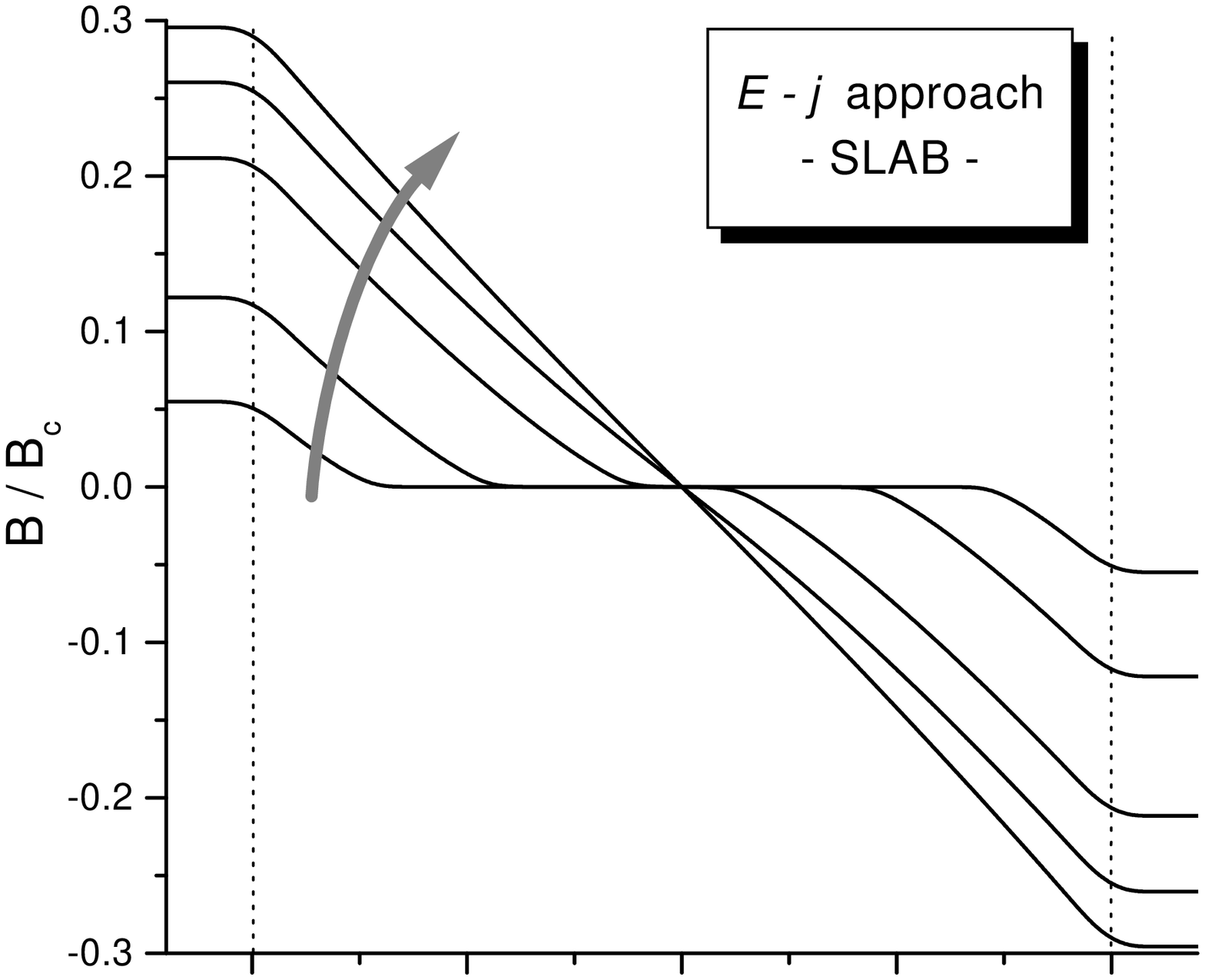,width=8cm}}
\centerline{ \psfig{figure=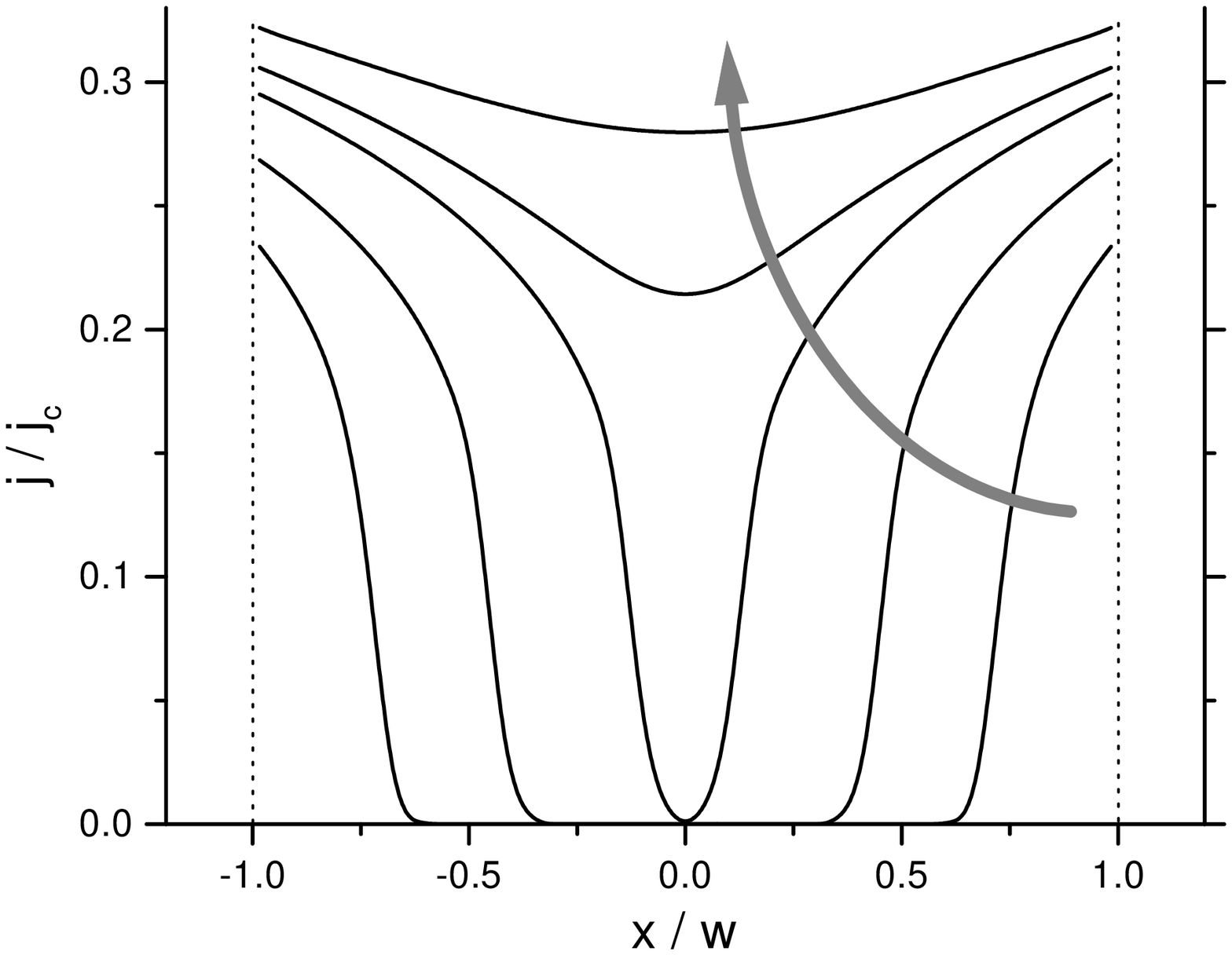,width=8cm}}
\caption{Same quantities as shown in \p\f{f_slb} only here calculated using
the $E-j$ approach, and evaluated for $\It/J_c=$ 0.06, 0.13, 0.21, 0.26, 0.30.
\label{f_slbb}}
\end{figure}

Figures~\ref{f_slb} and~\ref{f_slbb} present the time development of the
current and flux density distributions in the case of a slab.\cite{web}
In the flux penetrated region, which gradually expands from the edges,
both approaches lead to $B$-profiles that are essentially linear, and a fairly
constant current density. The slab has a central region where both $B$ and $j$
vanish. As $\It(t)$ increases, the flux penetrates deeper,
and the current becomes distributed more uniformly.
Although the overall behavior resembles that of the Bean model, one can also see
clear deviations, in particular in the current distributions.

The results also reveal distinct differences between the two approaches.
The most prominent one is seen in the $j$-distribution, where in the flux creep
approach a peak develops in the center as the slab becomes fully penetrated.
Another difference is visible near the edges, where the slopes in $j(x)$ are
significantly larger in the $E-j$ model. Both of these features are also reflected
in the $B$-distributions, although there only as different
curvatures of the profiles.

Subsequent increase in the transport current
(not shown in figures) does not change significantly the shape of
distributions. In the $E-j$ approach, the current density
tends to become completely uniform. In the flux creep approach,
the peak in the center retains, and both $j(x)$ and $B(x)$ increase
monotonously as grows the transport current, $\It$.

\subsection{Strip with a transport current}

\begin{figure}
\centerline{ \psfig{figure=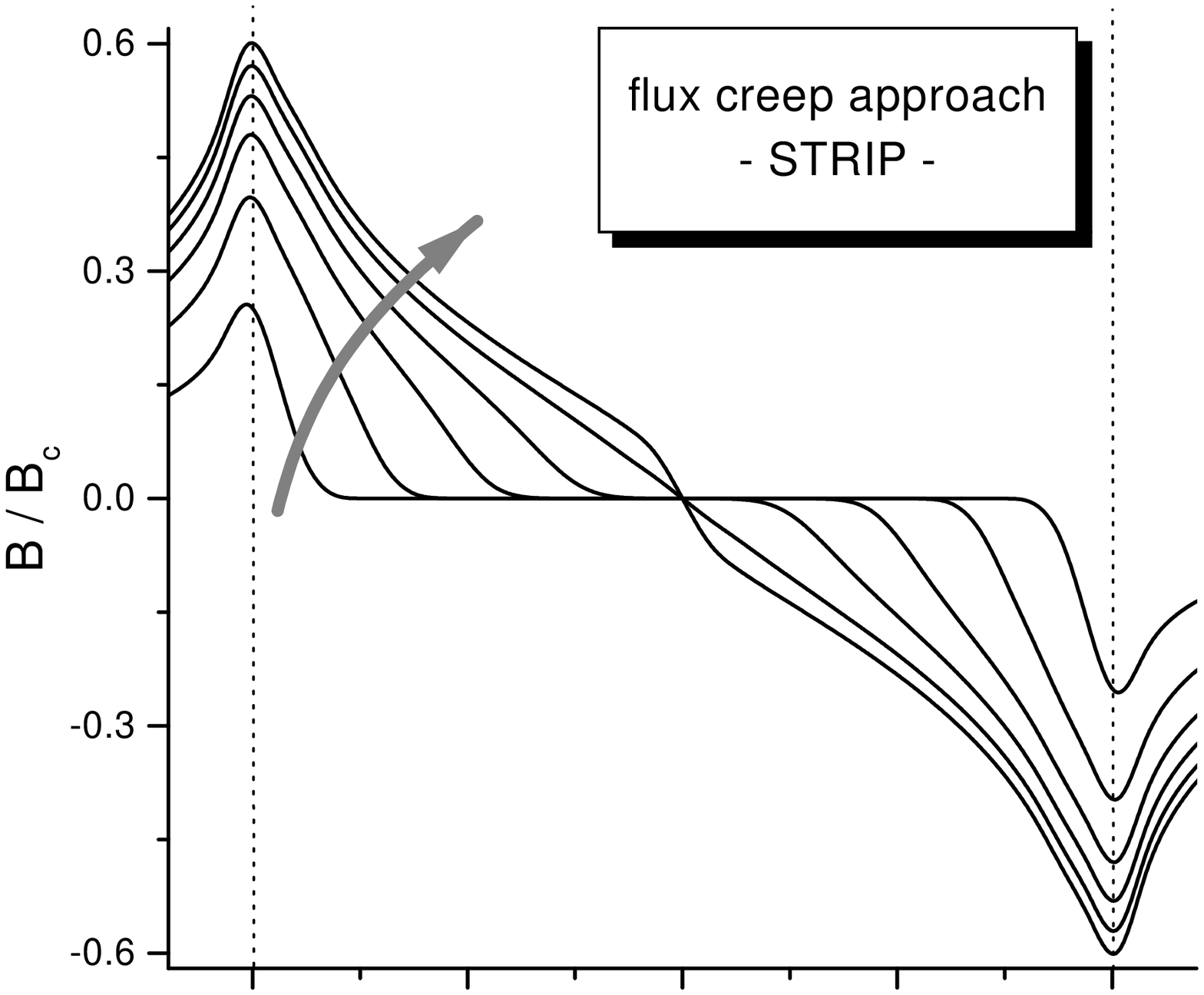,width=8cm}}
\centerline{ \psfig{figure=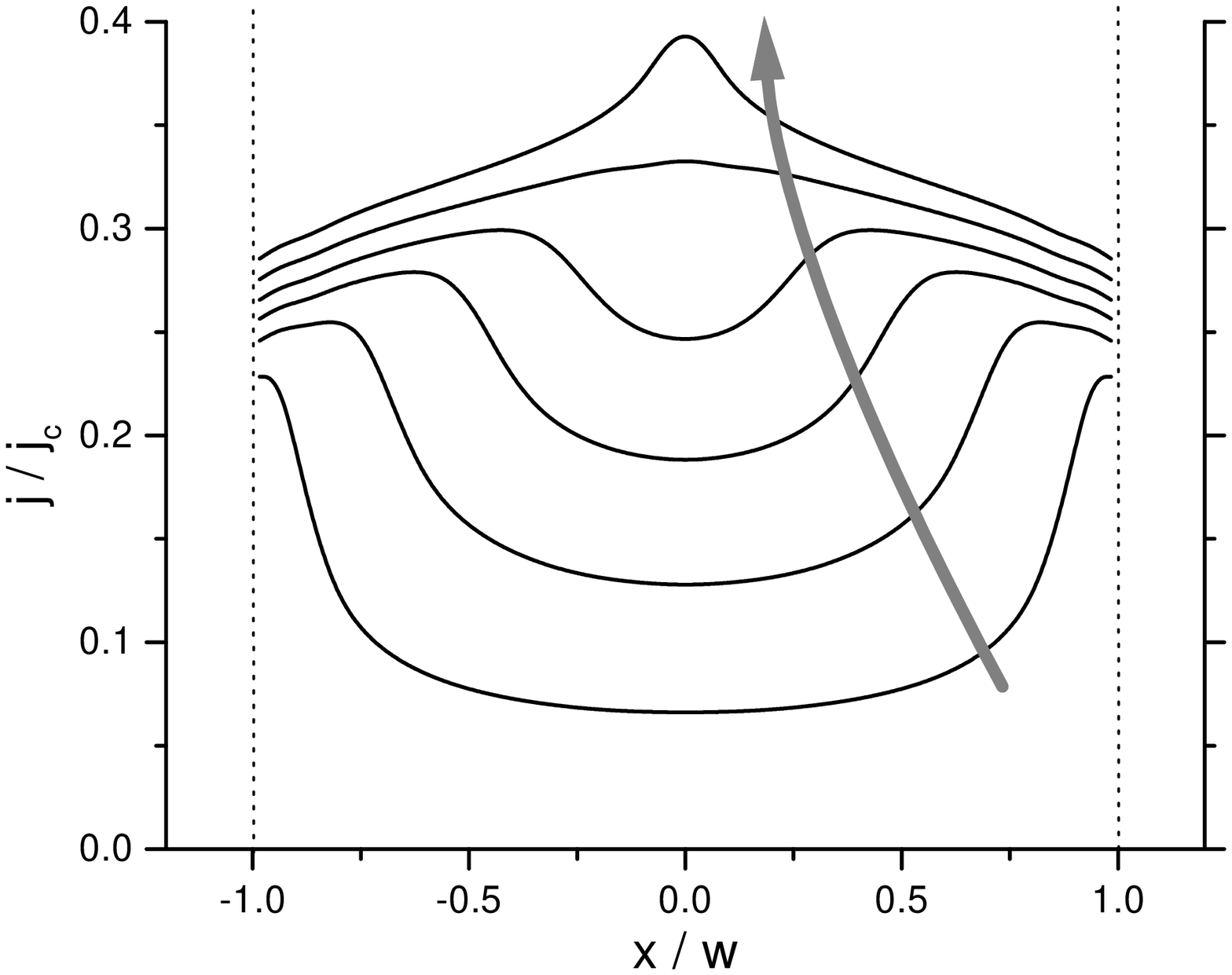,width=8cm}}
\caption{Temporal evolution of flux and current density distribution
in a thin strip with an increasing transport current.
The flux creep approach is used to calculate the graphs.
The profiles correspond to the current values
$I_T/I_c=$ 0.10, 0.18, 0.24, 0.28, 0.31, 0.33, where $I_c=2wd\ j_c$.
The dotted line marks the edge of the superconductor.
Arrows indicate the direction of time.
\label{f_str}}
\end{figure}

\begin{figure}
\centerline{ \psfig{figure=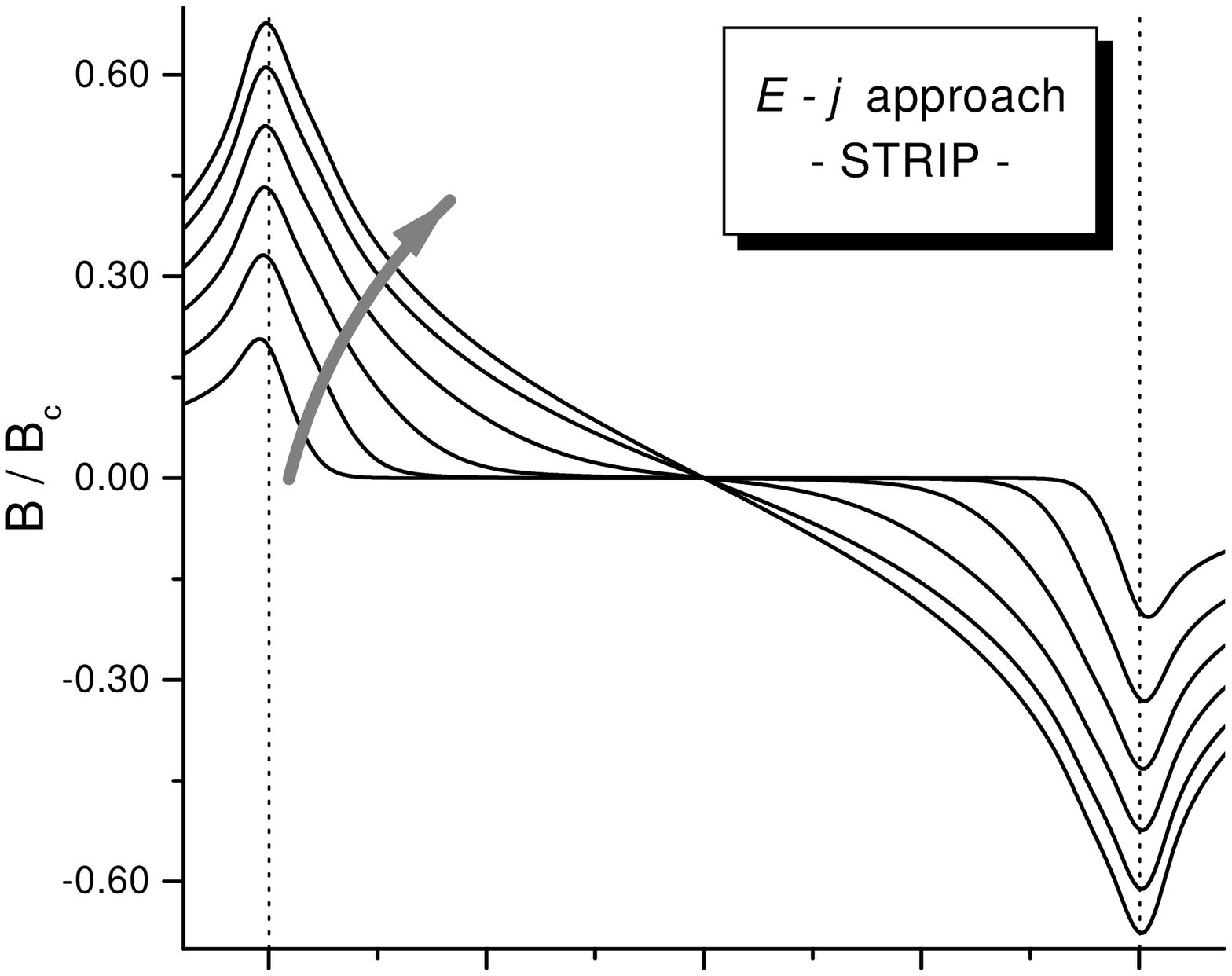,width=8cm}}
\centerline{ \psfig{figure=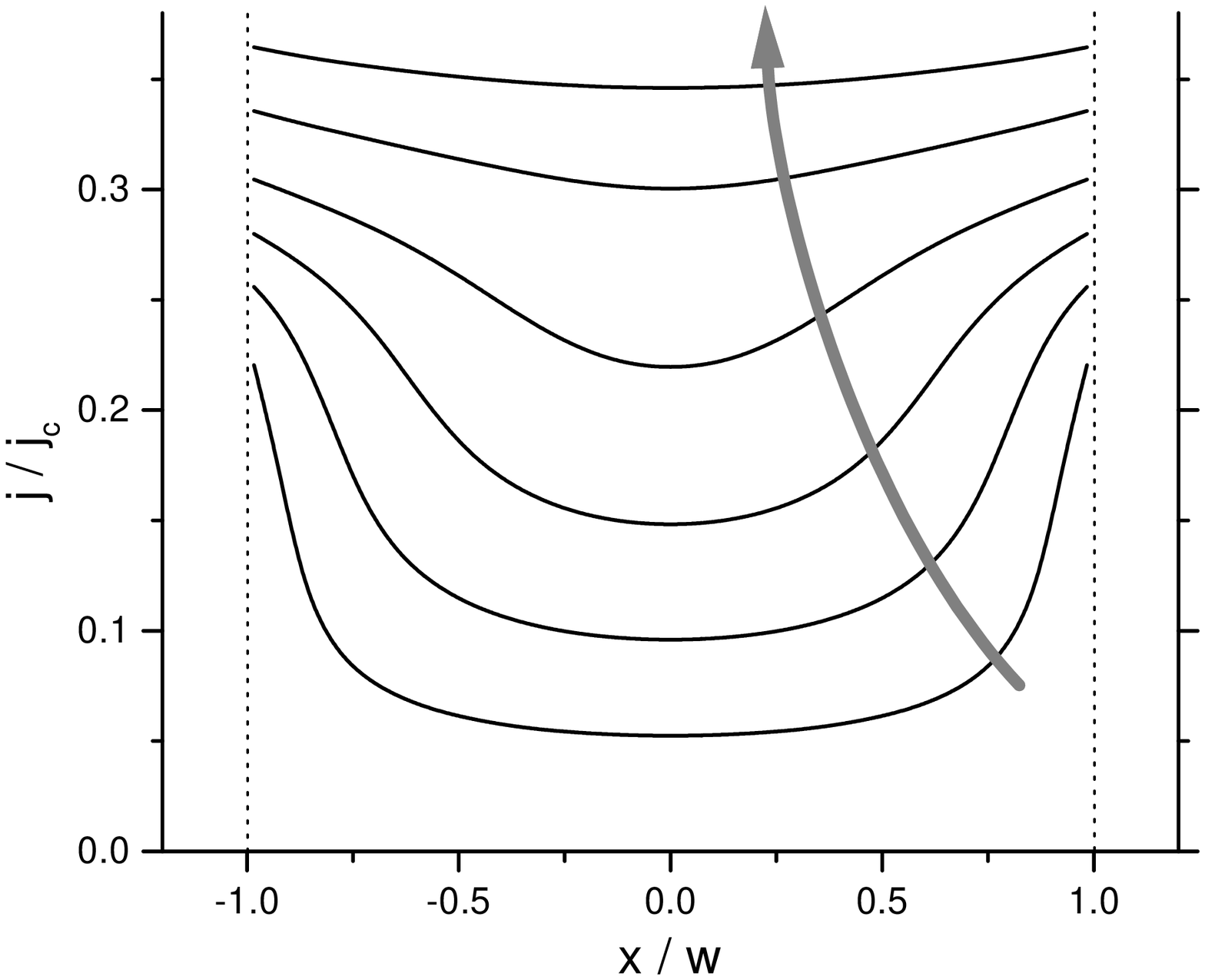,width=8cm}}
\caption{Same quantities as shown in \p\f{f_str} only here calculated using
the $E-j$ approach, and evaluated for $I_T/I_c=$ 0.08, 0.14, 0.20, 0.26,
0.32, 0.35.
\label{f_strb}}
\end{figure}

The simulated behavior of a thin strip experiencing a linearly
increasing transport current $I_T$, is shown in
Figures~\ref{f_str} and ~\ref{f_strb}.
The choise of parameters is the same as for the slab,
with $dI_T/dt = 10^{-3} j_c v_c d$
except
now it requires that $E_c= 0.2 v_c \mu_0 j_c d/\pi$, in order to give
approximately
the same average electric field for both the flux creep and the $E-j$ approach.
Comparing the results with the previous slab case, one immediatlely sees
differences in the shape of the profiles.
The $j(x)$ in a strip is always finite everywhere even at small currents, where the
flux penetration is only partial. Furthermore,
$B(x)$ is strongly nonlinear and has peaks at the edges. Both these features are well
known also in the Bean model behavior for a thin strip.\cite{BrIn,zeld}

As in the slab case, we observe also here a significant difference between
the distributions obtained from the flux creep and the $E-j$ approach.
Firstly, the two approaches give opposite sign for the slope of $j(x)$
in the penetrated regions near the edges.
Secondly, only the creep approach leads to a central peak in $j(x)$ at large
currents. Hence, while in the $E-j$ approach the $j(x)$ remains
concave throughout, the creep approach predicts a gradual change from a
concave to a convex profile. Contrary to the case of a slab, differences
are also clearly seen in the flux distributions. In
particular, the creep approach predicts a much steeper
slope near the flux front.

Interestingly, we found that although the two approaches lead to quite
different spatial distributions, the integral characteristics of the strip,
such as current-voltage curves, are only weakly sensitive to
the differences. This is demonstrated in \f{f_ivc}, which
shows the integral current-voltage curves obtained using both
approaches. The curves for $n=5$ correspond
to the $j$-distributions shown in the previous figures.
The electric field was determined as $P/I_T$, where the
dissipated power $P$ per unit length of the strip was
calculated by integrating the product $j(x)E(x)$
over the strip cross-section.\cite{BrIn}

In the log-log plot the current-voltage curves display a clear crossover
at $E \approx 0.002 E_c$.
At large currents the integral $E(j)$ curve shows a power law behavior
$E \propto j^{n'}$, where $n'$ is an ``integral'' exponent.
We find that $n'$ slightly exceeds $U_c/kT$ in the flux creep approach.
In the $E-j$ approach $n'$ is equal to
the ``local'' exponent $n$ and the integral $E(j)$ curves merge the
local $E(j)$'s shown by straight lines.
At small currents the integral $E(j)$
curve also shows a power law behavior, although with a much smaller $n'$
which seemingly does not depend on $n$ or $U_c/kT$.

\subsection{Discussion}

Both approaches describe the same physical situation: in response to
the transport current the flux lines enter the sample from the edge and
then move some distance before getting pinned or annihilated  in the
sample's center. Thus, near the edges the flux motion is always more
pronounced. Consequently, $E$ has a  maximum there. In the $E-j$ approach,
the local current density is an explicit function of the local electric
field. Therefore, $j(x)$ follows $E(x)$ and monotonously decreases
from the edges towards the center. On the other hand, in the flux creep
approach,
$j(x)$ is
related to $v(x)=E(x)/B(x)$ by Eq.~(\ref{E1}), and, hence
depends also on the flux distribution.
In particular, $j(x)$ is relatively small at the strip edges
where $|B|$ is maximal, see \f{f_str}(a).

\begin{figure}
\centerline{\psfig{figure=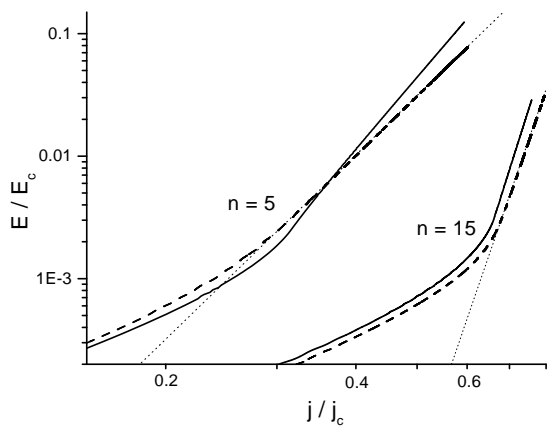,width=8cm}}
\caption{Current-voltage curves for a strip obtained by numerical
simulations in the flux creep approach (solid line)
and in the $E-j$ approach (dashed line). Results for two
different $n = U_c/kT$ are shown. Straight lines show the
local $E(j)$ curve for the $E-j$ approach, \p\eq{E2}.
\label{f_ivc}}
\end{figure}

\section{Experimental results and discussion}

A YBa$_2$Cu$_3$O$_{7-\delta}$ (YBCO) film of 200~nm thickness was prepared
by dc magnetron sputtering on LaAlO$_3$ substrate.\cite{kar1}
Using photo-lithography a strip of dimensions 500$\times$100~$\mu$m$^2$
was formed and equipped with Ag contact pads for injection of a
transport current. The current was applied in pulses of 40~ms duration
while the temperature was kept at 20~K in an optical cryostat.
Magneto-optical images were recorded with 33~ms exposure time
during the current pulse. From the images we
determine the $z$-component of $\bf B$
in the plane of the ferrite garnet magneto-optical indicator,
which we estimate to be located 10~$\mu$m above the
YBCO film.

Shown in \f{f_ex-j}(a) are the measured $B$-distributions.
Because of the finite distance between the indicator
and the superconductor, these profiles are not easily compared to
the results of the simulations. However, since the $j$-profiles
showed more distict differences between the creep and $E-j$ approach,
the measured $B$-profiles were converted to sheet current
distributions, $J(x)=d j(x)$, in the strip.
An inversion scheme described in Ref.~\onlinecite{joh96} and further
developed elsewhere\cite{future} was employed.

\begin{figure}
\centerline{ \psfig{figure=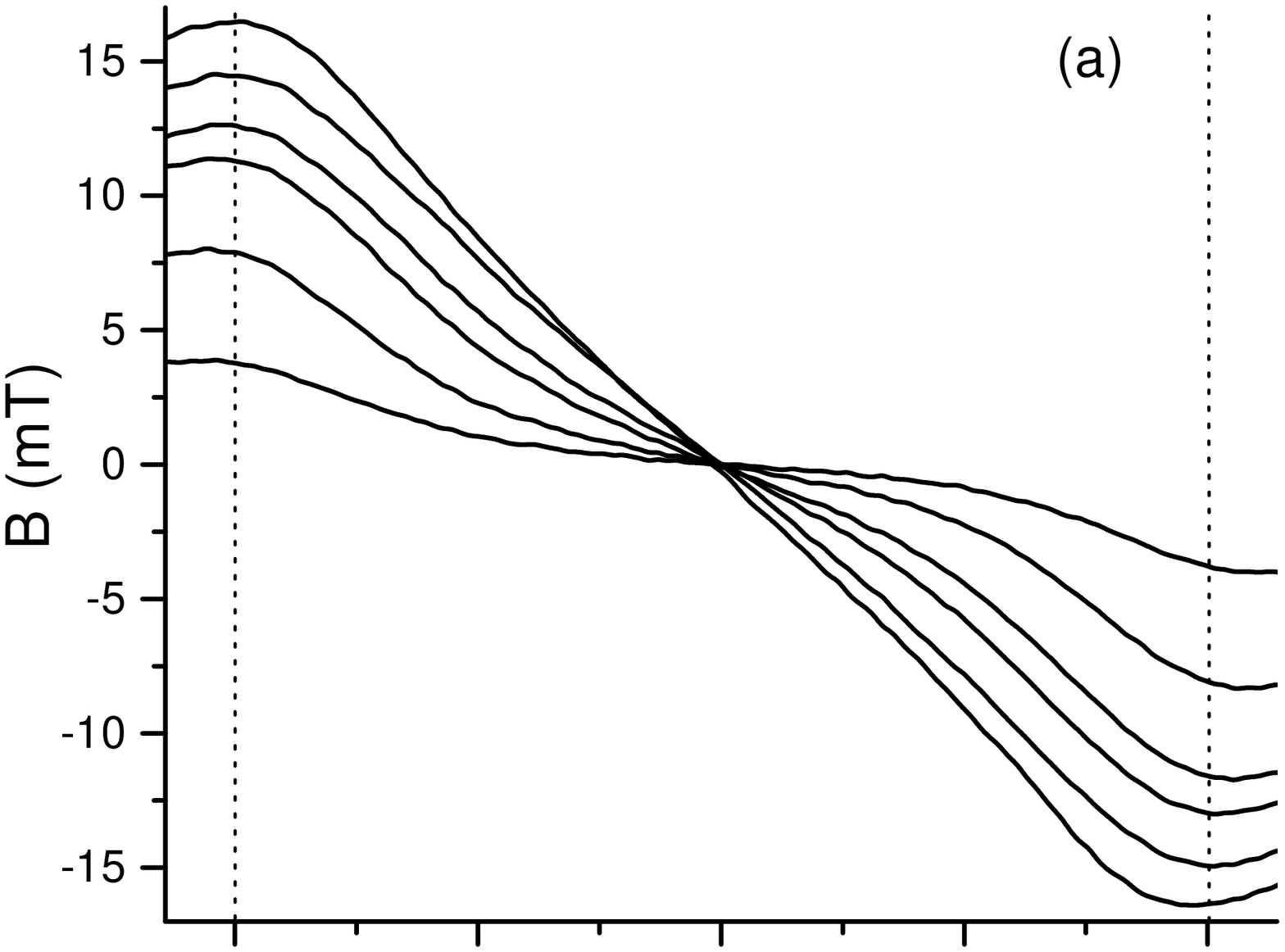,width=8cm}}
\centerline{ \psfig{figure=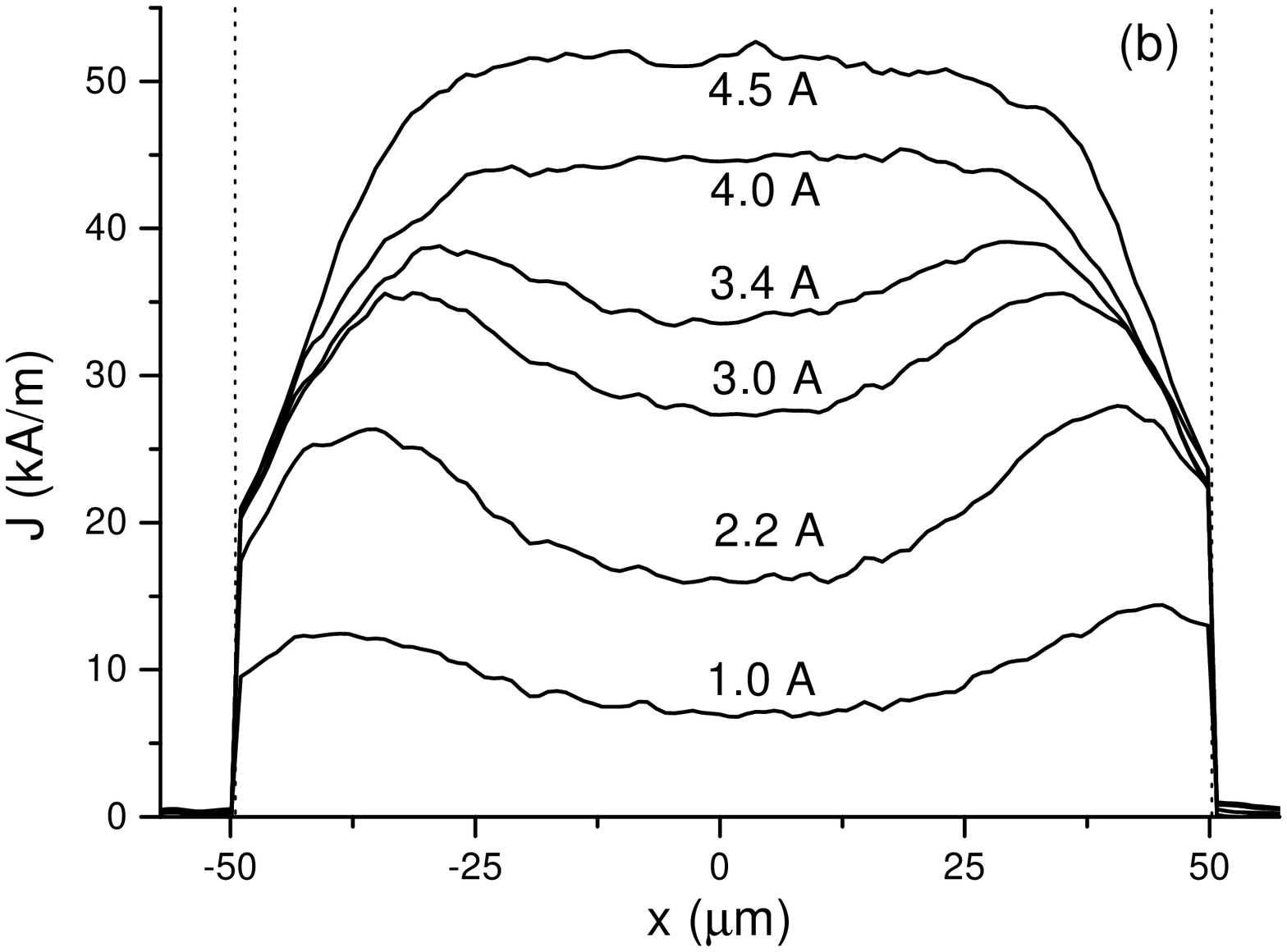,width=8cm}}
\caption{(a) Flux density distributions in a thin YBCO strip
carrying transport currents up to 4.5~A. The measurements
were obtained using magneto-optical imaging.
(b) Sheet current distributions inferred by inversion of the
$B(x)$ profiles.
\label{f_ex-j}}
\end{figure}

Profiles of the sheet current are shown in \f{f_ex-j}(b) for
a range of transport currents up to 4.5~A.
Evidently, they fit quite well to the simulated results
of the creep approach shown in \f{f_str}.
In particular, one easily recognizes the characteristic change from a
concave to a convex current distribution as $I_T$ increases.
The $E-j$ approach, on the other hand, appears not to be able to give
an adequate description of the flux dynamics in the present experiment.

Although being the better model, one can still see considerable
discrepancies between the experimental curves and the flux-creep
approach simulations. One example is
the peak of $j(x)$ in the strip center at large
currents, a feature the experiments could not reproduce.
Unfortunately, a current of 4.8~A caused fatal damage
to the sample, and we were not able to measure distributions under
the conditions where a central peak might become apparent.

\begin{figure}
\centerline{ \psfig{figure=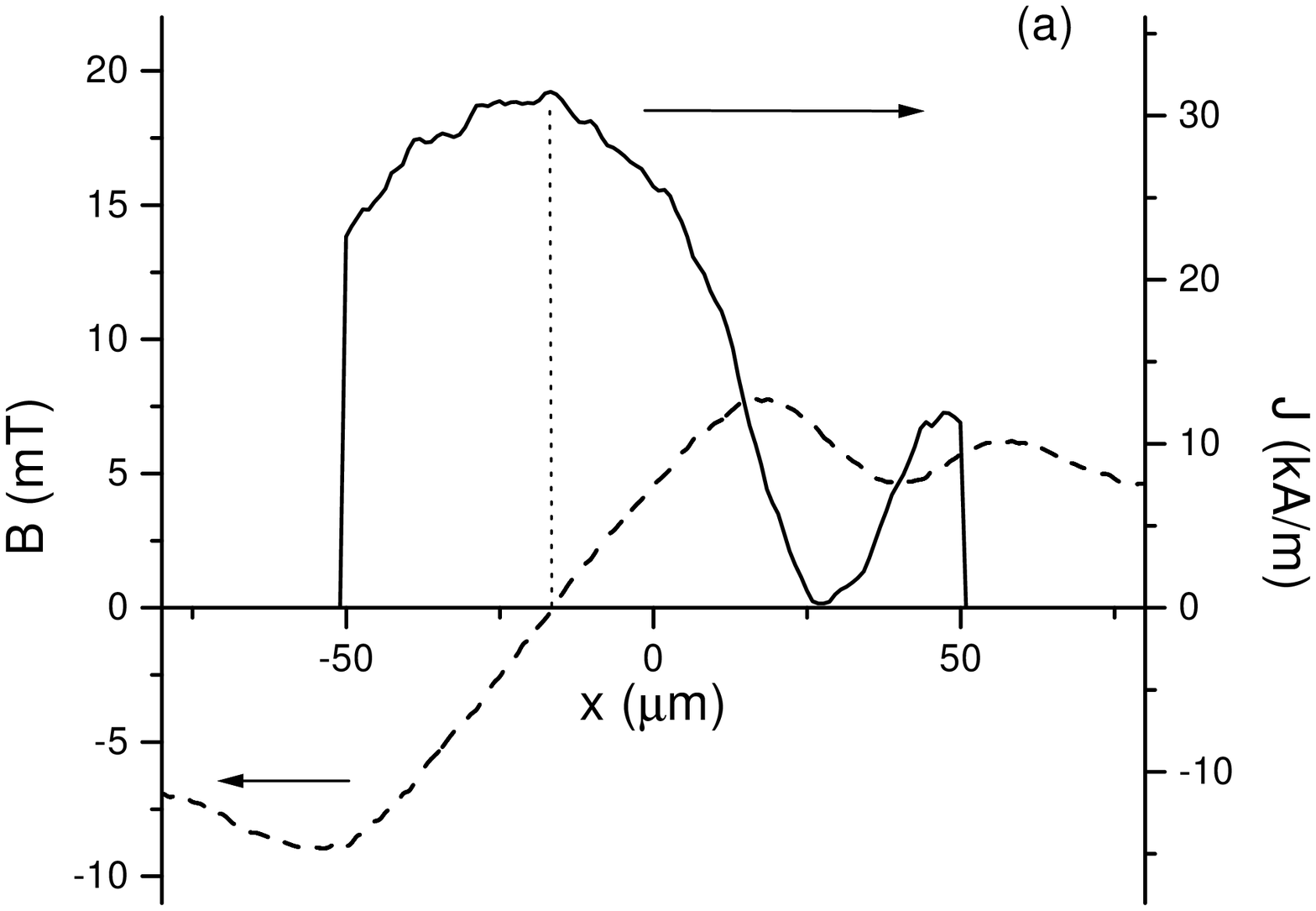,width=8cm}}
\centerline{ \psfig{figure=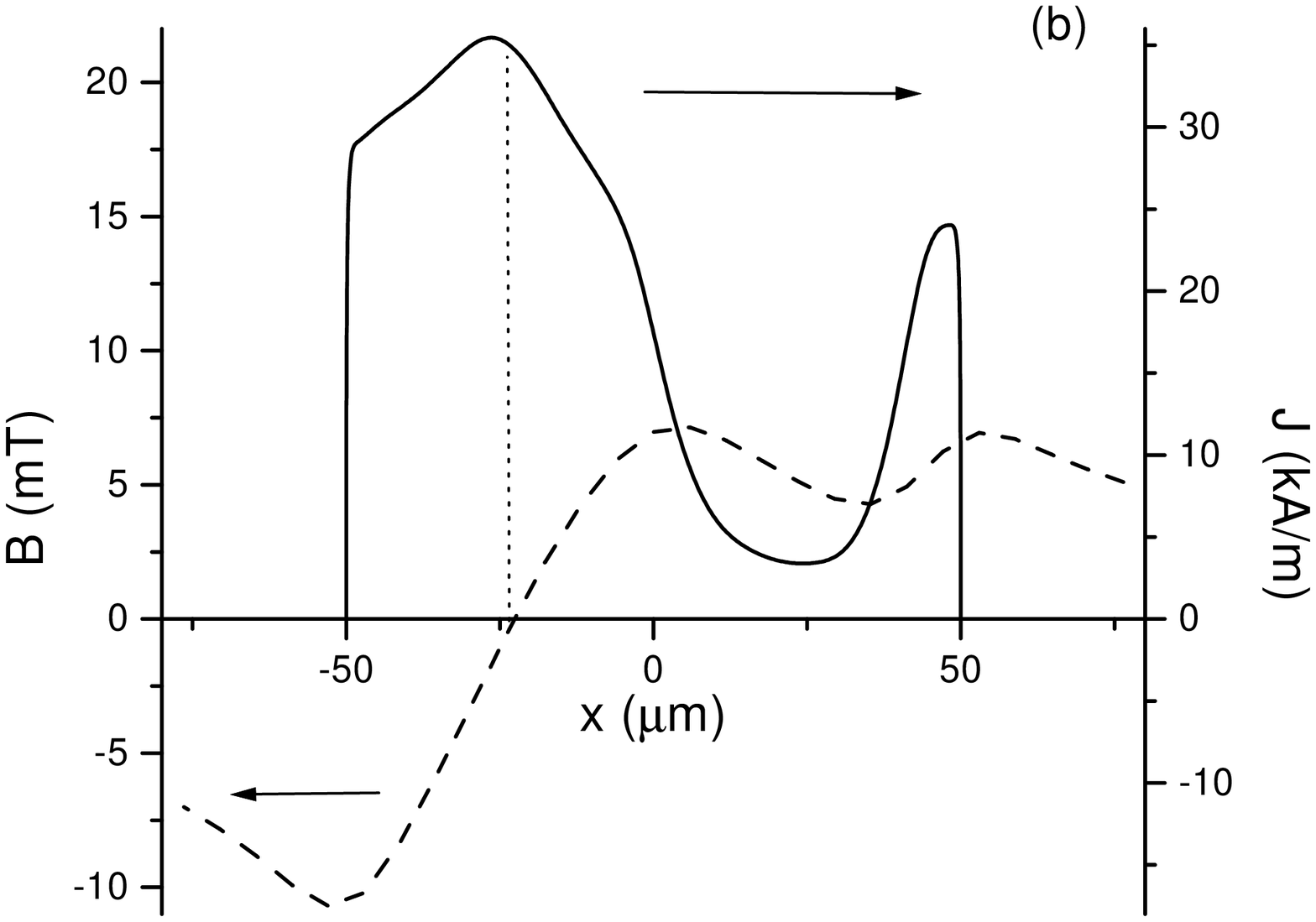,width=8cm}}
\centerline{ \psfig{figure=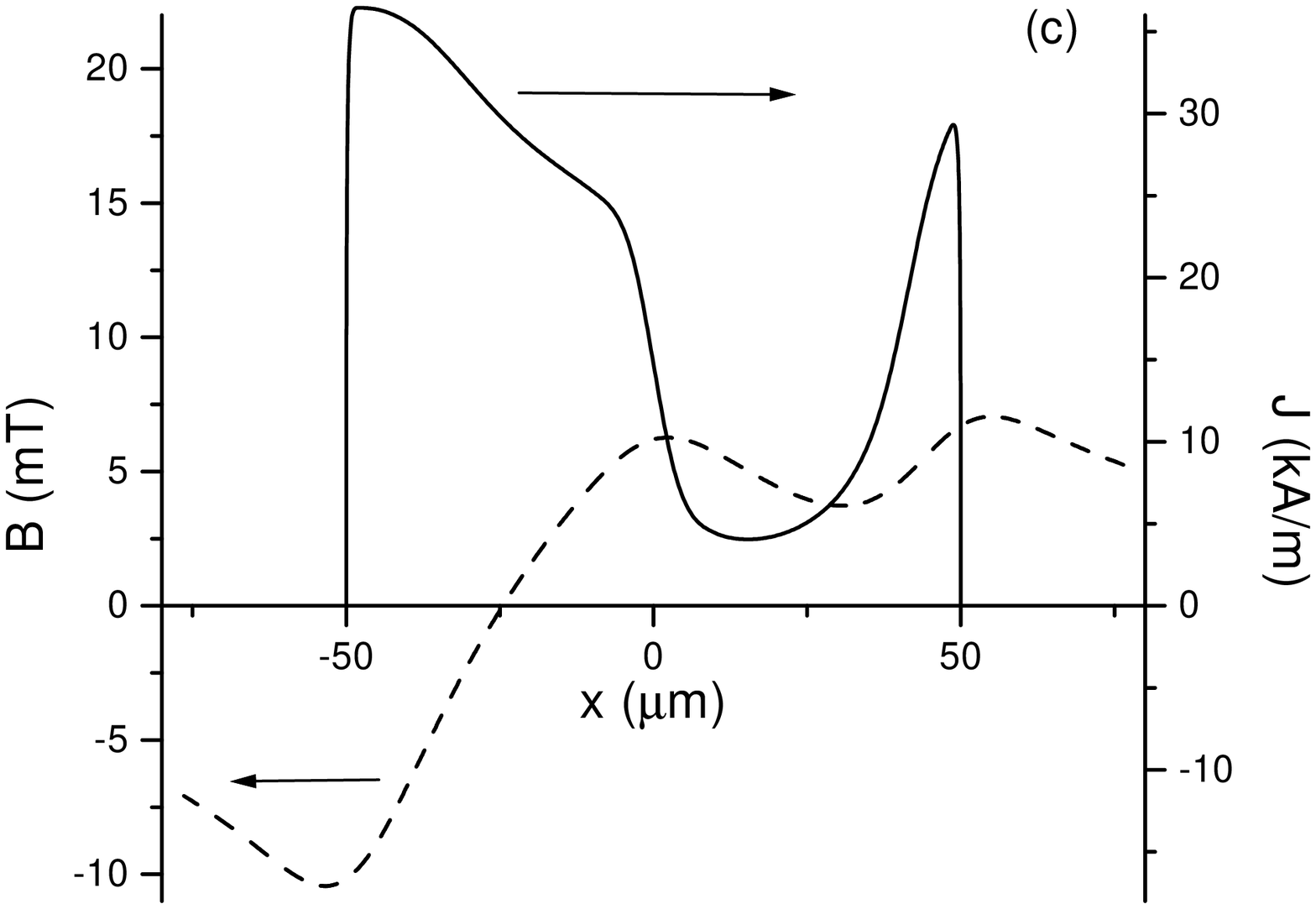,width=8cm}}
\caption{Sheet current (solid lines) and the flux density (dashed
lines) distributions in a thin strip: experiment (a) and simulations
in the flux creep approach (b) and simulations in the $E-j$
approach (c). The strip was first exposed to a very high magnetic
field which subsequently was reduced to zero. After that
a transport current of 2~A was applied. The experimental data
were obtained from magneto-optical images of a YBCO strip.
\label{f_rbj}}
\end{figure}
\noindent
As an alternative way to create a peak in $J$ where $B$ changes sign,
we carried out a different experiment. Here
the strip (a new YBCO sample prepared by the same method)
was initially in the remanent state after first being
exposed to a very high magnetic field.
After that the strip was subjected to a transport current of 2~A.
The resulting flux and current distributions are shown in \f{f_rbj}(a).
One sees that in the left half of the strip there is a wide region
with a large and nearly constant current density.
Within this region one finds that $J(x)$ indeed has a peak located
close to the point where $B=0$.

A new set of simulations was made for this
special state with combined magnetization currents and $I_T$.
The simulations aimed to reproduce the exact experimental steps:
First, the strip was exposed to a
perpendicular magnetic field applied as a pulse
of 70~mT amplitude with 1~s of linear increase and 1~s
of linear decay. Then, after one more second, the strip
was subjected to a transport current pulse with 1~ms 
rise time. 
During the current pulse, 
20~ms after turn-on, the magneto-optical image was taken.
For this particular time,
Figs.~\ref{f_rbj}(b) and (c) present the numerically obtained
distributions  for the flux creep approach and
$E-j$ approach, respectively.
 The following model parameters
were used: $n=U_c/kT=5$, $I_c=25$~A, and $v_c=10$~m/s.
Again, only the flux creep approach gives a peak in the current
profile, and now we find an excellent agreement with the experimental
results.

A remaining discrepancy between the flux-creep approach simulations
and the experimental curves is that the gradient of $J(x)$ near the
strip edge is larger in the experiments. This holds
true also for simulations made with other values of the power $U_c/kT$.
We believe that our photo-lithographic technology does not reduce the
film quality much more than up to a distance 1-2~$\mu$m from the edge.
This is also consistent with the magneto-optical images, which
show that the current flow along the strip is highly uniform
on  scales larger than 5~$\mu$m. Therefore, the discrepancy between
experiment and simulations indicates that the vortex behavior
is more complicated than assumed in the present flux creep model.
The experimentally observed suppression of $J$ near the edge
where $|B|$ is maximal, can be interpreted as  a $|B|$-induced
reduction of the critical current density $\jcu$,
or the pinning energy $U_c$. This interpretation, however,
fails to account for a similar
suppression of $J$ observed
previously in the remanent state after current pulse.\cite{mocur}
An alternative explanation able to cope with both
observations is a
heat dissipation due to vortex motion which is always most intensive
near the strip edges.

\section{Conclusions}

Numerical simulations have been carried out in order to
compare two commonly accepted approaches for analysis
of flux motion in superconductors;
(i) the flux creep approach,
and (ii) the approach based on a non-linear $E(j)$ curve.
We have shown that if the critical current density is
field-independent, these approaches predict similar but
also distinctly different current and flux distributions.
The difference is most pronounced in the regions where the
local flux density $B$ is small. The simulation results
were compared with the real current distributions
in a YBCO strip carrying a transport current. The experimental
data were obtained by using magneto-optical imaging. The comparison
shows clearly that the flux creep approach provides the better
description of the flux motion in the strip.

\acknowledgements

The financial support from the Research Council of Norway (NFR), and from
NATO via NFR is gratefully acknowledged.
We are grateful to Bj{\o}rn Berling for a 
many-sided help and to E. H. Brandt for a discussion.


\widetext

\begin{references}

\bibitem[*]{0}Email: t.h.johansen@fys.uio.no

\bibitem{giant} Y. Yeshurun and A. P. Malozemoff, Phys. Rev. Lett.
{\bf 60}, 2202 (1988).

\bibitem{yeshurun} Y. Yeshurun, A. P. Malozemoff, and A. Shaulov,
Rev. Mod. Phys. {\bf 68}, 911 (1996).


\bibitem{kes} H. G. Schnack, R. Griessen, J. G. Lensink, C. J. van der
Beek, and  P.~H.~Kes, Physica C {\bf 197}, 337 (1992).
\bibitem{burlachkov} L. Burlachkov, D. Giller, and R. Prozorov,
\prb {\bf 58}, 15067 (1998).


\bibitem{GurBra94} A. Gurevich and E. H. Brandt, Phys. Rev. Lett.  {\bf
73}, 178 (1994).
\bibitem{Br94} E. H. Brandt, Phys. Rev. B {\bf 49}, 9024 (1994).
\bibitem{Br-disk1} E. H. Brandt, \prb {\bf 58}, 6506 (1998).


\bibitem{equivalence}
The approaches would become equivalent for a special choice of the
parameters involved. In particular, if $E_c$ is proportional to $B$,
or if the critical current densities 
are related to each other as $\jce(B) = \jcu(B) B^{-1/n}$ with $n = U_c/kT$.

\bibitem{scal} Y. M. Galperin and T. H. Johansen, unpublished.

\bibitem{web} Movies of time development of $B(x)$, $j(x)$, and $E(x)$ 
distributions are presented at
http://www.fys.uio.no/faststoff/ltl/results/creep


\bibitem{BrIn}  E. H. Brandt, and M. Indenbom, Phys. Rev. B {\bf 48}, 12893
(1993).

\bibitem{zeld} E. Zeldov,  J. R. Clem, M. McElfresh,
and M. Darwin,  Phys. Rev. B {\bf 49}, 9802 (1994).

\bibitem{kar1} S. F. Karmanenko,  V. Y. Davydov,
M. V. Belousov,  R. A. Chakalov, G. O. Dzjuba, R. N. Il'in,
A. B. Kozyrev,  Y. V. Likholetov, K. F. Njakshev, I. T. Serenkov,
O. G. Vendic,  Supercond. Sci. Technol. {\bf 6}, 23 (1993).

\bibitem{joh96} T. H. Johansen,   M. Baziljevich,
 H. Bratsberg,  Y. Galperin, P. E. Lindelof, Y. Shen, and P. Vase,
 Phys. Rev. B {\bf  54}, 16 264 (1996).

\bibitem{future} A. V. Bobyl et al., unpublished.


\bibitem{mocur} M. E. Gaevski, A. V. Bobyl, D. V. Shantsev, S. F. Karmanenko, 
Y. M. Galperin, T. H. Johansen, M. Baziljevich, H. Bratsberg, 
Phys. Rev. B {\bf 59}, 9655 (1999)



\end{references}
\end{document}